\documentclass[preprint,aps,tightenlines,eqsecnum,floatfix,showpacs]{revtex4}
\usepackage{amsmath}
\usepackage{epsfig}
\usepackage{bm}
\usepackage{slashed}
\begin{document}
\title{Relativity constraints on the two-nucleon contact interaction}
\author{L.\ Girlanda$^{\,{\rm a,b}}$, S.\ Pastore$^{\,{\rm c}}$,
  R.\ Schiavilla$^{\,{\rm c,d}}$, and
M.\ Viviani$^{\,{\rm b}}$}
\affiliation{
$^{\,{\rm a}}$\mbox{Department of Physics, University of Pisa, 56127 Pisa, Italy}\\
$^{\,{\rm b}}$\mbox{INFN-Pisa, 56127 Pisa, Italy}\\
$^{\rm c}$\mbox{Department of Physics, Old Dominion University, Norfolk, VA 23529, USA}\\
$^{\rm d}$\mbox{Jefferson Lab, Newport News, VA 23606, USA}\\
}
\date{\today}
\begin{abstract}
We  construct the most general, relativistically invariant, contact Lagrangian
at order $Q^2$ in the power counting, $Q$ denoting the low momentum scale.
A complete, but non-minimal, set of (contact) interaction terms is identified, which
upon non-relativistic reduction generate 2 leading independent operator combinations of
order $Q^0$ and 7 sub-leading ones of order $Q^2$---a result derived previously in the
heavy-baryon formulation of effective field theories (EFT's).  We show that Poincar\'e covariance
of the theory requires that additional terms with fixed coefficients be included, in order
to describe the two-nucleon potential in reference frames other than the center-of-mass
frame.  These terms will contribute in systems with mass number $A>2$, and their
impact on EFT calculations of binding energies and scattering observables in these
systems should be studied.
\end{abstract}

\maketitle

\section{Introduction and Conclusions}

Chiral effective field theory ($\chi$EFT), pioneered by Weinberg in
a series of seminal papers almost two decades ago~\cite{Weinberg90},
has led to a novel understanding of strong interactions in nuclei by
providing a direct link between these interactions and
the symmetries of quantum chromodynamics, including chiral
symmetry with its explicit and dynamical breaking mechanisms (see review papers in
Refs.~\cite{Bernard95}).  In its original form, $\chi$EFT is formulated in
terms of pions and (non-relativistic) nucleons, whose interactions, strongly constrained by
chiral symmetry, are organized as an expansion in powers of small
momenta $Q$.  All heavier degrees of freedom are ``integrated out'',
and their effects are implicitly subsumed in the coupling constants
accompanying local vertices.  At sufficiently low energy, even
the pions can be integrated out, and the nucleons only interact through
contact vertices.  In either case, two-nucleon ($NN$) contact interactions
are an important aspect of EFT descriptions.  In the
present paper we examine the constraints that relativistic covariance
imposes on the resulting $NN$ potential up to order $Q^2$ (or next-to-next
leading order, N$^2$LO).

At LO ($Q^0$) in the low energy expansion there are only 2
independent contact interactions~\cite{Weinberg90}
\begin{equation} 
\label{eq:op0lagr}
{\cal L}_I^{(0)}=-\frac{1}{2} C_S\, O_S -\frac{1}{2} C_T\, O_T \ ,
\end{equation}
where $C_S$ and $C_T$ denote low-energy constants (LEC's), and
the operators $O_S$ and $O_T$ are defined in terms of
the non-relativistic nucleon field $N(x)$,
\begin{equation}
N(x) = \int \frac{d{\bf p}}{(2\pi)^3}\, b_s({\bf  p}) \, \chi_s\, {\rm e}^{-i p \cdot x} \ ,
\label{eq:nrf}
\end{equation}
and its adjoint in Table~\ref{tb:tab1}.  Here $b_s({\bf p})$ and $b^\dagger_s({\bf p})$
are annihilation and creation operators for a nucleon in spin state $s$,
satisfying standard anticommutation relations, {\it i.e.}
$\left[b_s({\bf p})\, , \, b^\dagger_{s^\prime }({\bf p}^\prime)\right]_+ = (2\pi)^3\delta({\bf p}-{\bf p}^\prime)\,
\delta_{ss^\prime}$.  A sum over the repeated index
$s=\pm1/2$ is implied, and it is understood that field operator
products are normal-ordered in $O_S$ and $O_T$ (as well as in the $O_i$'s defined below).
We have suppressed isospin indices, since they will
not enter in the discussion to follow (see Sec.~\ref{sec:nr1}).  The corresponding $NN$ potential
reads
\begin{equation}
v^{\rm CT0}=C_S+C_T\,{\bm \sigma}_1\cdot{\bm \sigma}_2 \  .
\label{eq:ct0}
\end{equation}
\begin{center}
\begin{table}[bth]
\begin{tabular}{c|c}
\hline
\hline
$O_S$  & $(N^\dagger N)(N^\dagger N)$ \\
$O_T$  & $(N^\dagger {\bm {\sigma}} N)\cdot 
(N^\dagger{\bm {\sigma}}N)$ \\
\hline
$O_1$  & $(N^\dagger \overrightarrow{\bm \nabla} N)^2 
+{\rm h.c.}$    \\
$O_2$  & $(N^\dagger \overrightarrow{\bm \nabla} N )\cdot
( N^\dagger \overleftarrow{\bm \nabla} N) $\\
$O_3$  &  $(N^\dagger N) ( N^\dagger \overrightarrow{\bm \nabla}^2 N)+{\rm h.c.}$  \\
$O_4$  & $i \,( N^\dagger \overrightarrow{\bm \nabla} N) \cdot (N^\dagger \overleftarrow{\bm \nabla}
  \times {\bm \sigma} N )+ {\rm h.c.}$ \\
$O_5$ & $i \, (N^\dagger N)(N^\dagger \overleftarrow{\bm \nabla}
\cdot {\bm \sigma} \times \overrightarrow{\bm \nabla} N)$ \\
$O_6$ & $i \, (N^\dagger {\bm \sigma} N) \cdot
(N^\dagger \overleftarrow{\bm \nabla} \times 
\overrightarrow{\bm \nabla} N)$\\
$O_7$  & $( N^\dagger {\bm \sigma} \cdot
\overrightarrow{\bm \nabla} N) (N^\dagger {\bm \sigma}\cdot \overrightarrow{\bm \nabla} N) +{\rm h.c.}$ \\
$O_8$  & $(N^\dagger \sigma^j
\overrightarrow{\nabla^k} N)(N^\dagger \sigma^k \overrightarrow{\nabla^j} N) + {\rm h.c.}$  \\
$O_9$  &   $(N^\dagger \sigma^j
\overrightarrow{\nabla^k} N)(N^\dagger \sigma^j \overrightarrow{\nabla^k} N) + {\rm h.c.}$ \\
$O_{10}$  & $(N^\dagger {\bm \sigma} \cdot
\overrightarrow{\bm \nabla}N) 
(N^\dagger \overleftarrow{\bm \nabla}\cdot {\bm \sigma} N)$ \\ 
$O_{11}$  & $(N^\dagger \sigma^j \overrightarrow{\nabla^k} N) 
(N^\dagger \overleftarrow{\nabla^j} \sigma^k  N)$ \\
$O_{12}$  &  $(N^\dagger \sigma^j \overrightarrow{\nabla^k} N) 
(N^\dagger \overleftarrow{\nabla^k} \sigma^j  N)$\\
$O_{13}$  &  $ (N^\dagger
\overleftarrow{\bm \nabla}\cdot{\bm  \sigma} \,\overrightarrow{\nabla^j} N)
(N^\dagger \sigma^j N) +{\rm h.c.}$\\
$O_{14}$  & $2\, (N^\dagger
\overleftarrow{\bm \nabla} \sigma^j \cdot \overrightarrow{\bm \nabla} N) (N^\dagger \sigma^j N)$ \\
\hline
\hline
\end{tabular}
\caption{Operators entering the LO ($Q^0$) and N$^2$LO ($Q^2$) contact
interactions~\protect{\cite{Ordonez96}}.  The left (right) arrow on $\nabla$  indicates that the gradient  acts on the left (right) field.
Normal-ordering of the field operator products is understood.}
\label{tb:tab1}
\end{table}
\end{center}

At the next non-vanishing order, N$^2$LO, the contact Lagrangian
involving two derivatives of the nucleon fields has been written in
Ref.~\cite{Ordonez96} as consisting of 14 operators
\begin{equation}
\label{eq:op2lagr}
{\cal L}_I^{(2)} = - \sum_{i=1}^{14} C^\prime_i \, O_i \ ,
\end{equation}
where the $O_i$'s are listed in Table~\ref{tb:tab1} and the
$C_i^\prime$ are LEC's.  In fact, we showed in Ref.~\cite{Pastore09}
that, after partial integrations, only 12 out of the
above 14 operators are independent, since
\begin{equation}
O_7 + 2\,  O_{10} = O_8 + 2\, O_{11}\ , \quad O_4 +O_5 = O_6 \ .
\end{equation}
In a general frame in which the $NN$ pair has total momentum ${\bf P}$ and
initial and final relative momenta, respectively, ${\bf p}$ and ${\bf p}^\prime$,
the potential derived from the Lagrangian ${\cal L}_I^{(2)}$ is conveniently
separated into a term, $v^{\rm CT2}$, independent of ${\bf P}$~\cite{Ordonez96,Epelbaum98}
and one, $v_{\bf P}^{\rm CT2}$, dependent on it~\cite{Pastore09}:
\begin{eqnarray}
v^{\rm CT2}({\bf k}, {\bf K})&=& C_1\,k^2+C_2\,K^2+
(C_3\,k^2+C_4\,K^2)\,{\bm \sigma}_1\cdot{\bm \sigma}_2
+ i\,C_5\,\frac{{\bm \sigma}_1+{\bm \sigma}_2}{2}\cdot {\bf K}\times{\bf k} \nonumber\\
&+&C_6\,{\bm \sigma}_1\cdot{\bf k}\,\,{\bm \sigma}_2\cdot {\bf k}
+C_7\,{\bm \sigma}_1\cdot{\bf K}\,\,{\bm \sigma}_2\cdot {\bf K} \ ,
\label{eq:ct2} 
\end{eqnarray}
\begin{eqnarray}
v_{\bf P}^{\rm CT2}({\bf k},{\bf K})&=&
i\,C^*_1\,\frac{{\bm \sigma}_1-{\bm \sigma}_2}{2}\cdot {\bf P}\times{\bf k}
+C^*_2\,({\bm \sigma}_1\cdot{\bf P}\,\,{\bm \sigma}_2\cdot {\bf K}
-{\bm \sigma}_1\cdot{\bf K}\,\,{\bm \sigma}_2\cdot {\bf P}) \nonumber \\
&+&(C^*_3+C^*_4\,{\bm \sigma}_1\cdot{\bm \sigma}_2)\, P^2
+C^*_5\,{\bm \sigma}_1\cdot{\bf P}\,\,{\bm \sigma}_2\cdot {\bf P} \ ,
\label{eq:ct2p}
\end{eqnarray}
where the momenta ${\bf k}$ and ${\bf K}$ are defined as
${\bf k} = {\bf p}^{\prime} -{\bf p}$ and ${\bf K}=({\bf p}^{\prime}+{\bf p})/2$, and
the $C_i$'s ($i=1,\dots,7$) and $C^*_i$ ($i=1,\dots,5$)  are in a one-to-one correspondence
with the LEC's $C_i^{\prime}$'s multiplying the 12 independent operators
(see Refs.~\cite{Ordonez96,Pastore09,Epelbaum98}).

We argued in Ref.~\cite{Pastore09} that the ${\bf P}$-dependent terms
represent boost corrections to the LO potential $v^{\rm CT0}$, and that
the $C_i^*$, rather than being independent LEC's, are in fact related to
$C_S$ and $C_T$ as
\begin{equation}
C_1^*=\frac{C_S-C_T}{4 m^2}\ , \quad C_2^*=\frac{C_T}{2 m^2}\ , \quad
C_3^*=-\frac{ C_S}{4 m^2}\ , \quad C_4^*=-\frac{ C_T}{4 m^2}\ , \quad
C_5^*=0\ ,
\label{eq:cip}
\end{equation}
where $m$ is the nucleon mass.  This result is derived in
relativistic quantum mechanics---its instant-form
formulation~\cite{Dirac49}---by requiring that the commutation relations
of the Poincar\'e group generators are satisfied, which, to
order ${\bf P}^2/m^2$, leads to the elegant relation~\cite{Krajcik74,Carlson93}
\begin{equation}
 v_{\bf P}^{\rm CT2}=
-\frac{P^2}{8\, m^2} v^{\rm CT0} +\frac{i}{8\, m^2}\left[{\bf P}\cdot{\bf r}~
{\bf P}\cdot {\bf p}\, ,\,  v^{\rm CT0}\right]+\frac{i}{8\, m^2}
\left[({\bm \sigma}_1-{\bm \sigma}_2)\times{\bf P}
\cdot{\bf p}\, ,\, v^{\rm CT0}\right] \ ,
\label{eq:vcomm}
\end{equation}
where ${\bf r}$ and ${\bf p}$ are, respectively, the relative position and momentum
operators.  The potential $v_{\bf P}^{\rm CT2}({\bf k},{\bf K})$ then follows by
evaluating the commutators in momentum space, and by retaining only contributions
of order $Q^2$ (we assume here $P \sim k \sim K \sim Q$).  That there are dynamical
corrections to the $NN$ interaction, when it is boosted to an arbitrary frame, is not
surprising, as in instant-form relativistic quantum mechanics interactions enter
both the Hamiltonian and boost generators.

In the present paper, we justify the claim made above in a EFT
setting.  We proceed in two steps.  First, we construct, up to order
$Q^2$ included, the most general hermitian Lagrangian density allowed by
invariance under transformations of the Lorentz group and by the discrete
symmetries of the strong interaction.  After performing its non-relativistic
reduction, we find that there are 2 independent operator combinations
of order $Q^0$, accompanied by specific $Q^2$ corrections, and 7
independent operator combinations of order $Q^2$---a
result also obtained~\cite{Epelbaum00} in the heavy baryon
formulation~\cite{Georgi90} of ${\cal L}_I^{(2)}$ by
requiring that it be re-parametrization invariant~\cite{Luke92}.

Second, we show that this same picture emerges within
the non-relativistic theory in the context of a systematic
power counting, by enforcing that the commutation
relations among the Poincar\'e group generators are
satisfied order by order in the low energy expansion (for a similar
approach, in a different context, see Ref.~\cite{vairo}).
The above correspondence between the $C_i^*$'s and
$C_S$, $C_T$ is recovered, showing that the commutator
relation in Eq.~(\ref{eq:vcomm}) remains valid in a EFT framework.
Thus, in order to determine the boost corrections of order $Q^2$ to the
complete LO chiral potential, which also includes the one-pion-exchange
term, one can either use Eq.~(\ref{eq:vcomm}) or
compute the potential in an arbitrary frame
starting from the Lagrangian of the covariant theory.

These boost corrections should be taken into account in $\chi$EFT
(and EFT) calculations of nuclei
with mass number $A > 2$.  So far, they have been evaluated, for the case of
realistic potentials, in the $A$=3 and 4 binding energies, where they have
been found to give, respectively, about 400 keV and 1.9 MeV repulsive contributions~\cite{Carlson93},
as well as in three-nucleon scattering observables~\cite{Witala08}, where, in particular,
they have led to an increase of the discrepancy between theory and experiment
in the $nd$ vector analyzing power.

\section{Non-relativistic reduction}
\label{sec:nr}
\begin{figure}
\centerline{\includegraphics[width=7cm]{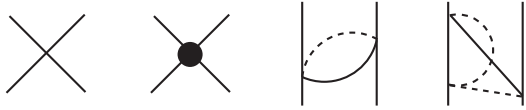}}
\caption{Time ordered diagrams contributing to the $NN$ scattering
amplitude and involving nucleons (solid lines) and
antinucleons (dashed lines) interacting through the contact
vertices at order $Q^0$ and $Q^2$ (solid circle).  Note that
at order $Q^2$ the diagrams with antinucleons do not contribute
(see text for explanation).}
\label{fig:nndiag}
\end{figure}

To begin with, we observe that, while the relativistic theory is written
in terms of fermion fields $\psi=\psi^{(+)} + \psi^{(-)}$ containing
both positive- and negative-energy components, the latter play no role
in the $NN$ contact potential at order $Q^2$ of interest here.  This
is because antinucleon degrees of freedom only enter via loop corrections,
and each loop is suppressed by a factor $Q^3$ in time ordered perturbation
theory (examples are shown in Fig.~\ref{fig:nndiag}).
Hence, the two-derivative contact Lagrangian (of order $Q^2$) can be derived,
without any loss of generality, starting from the relativistic theory
and ignoring the negative energy components.

\subsection{Generalities and strategy}
\label{sec:nr1}

The building blocks of the relativistic contact Lagrangian are
products of fermion bilinears with space-time structures
\begin{equation}
\label{eq:bb}
\frac{1}{ (2m)^{N_d}} (\overline{\psi}\,
i\!\overleftrightarrow{\partial}^\alpha \,
i\!\overleftrightarrow{\partial}^\beta  
\cdots \Gamma_A \, \psi )\, 
\partial^\lambda \, \partial^\mu \cdots (\overline{\psi}\,
i\!\overleftrightarrow{\partial}^\sigma \,
i\!\overleftrightarrow{\partial}^\tau \cdots  \Gamma_B\,   \psi )\ , 
\end{equation}
where 
$\overleftrightarrow{\partial}^\alpha = \overrightarrow{\partial}^\alpha -
\overleftarrow{\partial}^\alpha$ and the $\Gamma$'s denote
generic elements of the Clifford algebra, expanded in the usual
basis 1, $\gamma_5$, $\gamma^\mu$, $\gamma^\mu \gamma_5$,
$\sigma^{\mu \nu}$, or the Levi-Civita tensor $\epsilon^{\mu\nu\rho\sigma}$
(with the convention $\epsilon^{0123}=-1$). In the above equation,
$N_d$ stands for the number of four-gradients (both
$\overleftrightarrow{\partial}$ and $\partial^\lambda$) entering the formula,
and the factor ${1/(2m)^{N_d}}$ has been introduced so that all contact
terms will have the same dimension.
The Lorentz indices $\alpha,\dots, \tau$ on the partial
derivatives are contracted among themselves and/or with
those in the $\Gamma_{A,B}$ (for ease of presentation,
these indices, unless necessary, will be
suppressed hereafter).  In order to have flavor singlets,
the isospin structure of the two bilinears must be either
$1 \otimes 1$ or  $\tau^a \otimes \tau^a$.  However, the latter
needs not be considered, as it can be eliminated by Fierz
rearrangement.

A few remarks are now in order.  First, the Lagrangian density should
be hermitian and invariant under charge conjugation (${\cal C}$) and
parity (${\cal P}$).  We list the transformation properties
of the fermion bilinears in Table~\ref{tb:discrete}.
\begin{table}[bth]
\begin{tabular}{l|c|c|c|c|c|c|c|c}
\hline
\hline
 & 1& $\gamma_5$ & $ \gamma_\mu$ & $\gamma_\mu \gamma_5$ & $
 \sigma_{\mu \nu}$ & $ \epsilon_{\mu \nu
 \rho \sigma}$ & $\overleftrightarrow{\partial_\mu}$ & $\partial_\mu$
 \\
\hline
${\cal P}$ & + & -- & + & -- & + &  -- & + & + \\
\hline
${\cal C}$ & + & + & -- & + & -- & + & -- & +\\
\hline
h.c.\ & + & -- & + & + & + & + & -- & + \\
\hline
\hline
\end{tabular}
\caption{Transformation properties of the fermion bilinears
(with the different elements of the Clifford algebra), Levi-Civita tensor, and derivative operators
under parity (${\cal P}$), charge conjugation (${\cal C}$), and hermitian conjugation
(h.c.). The symbol $\partial_\mu$ in the last column stands for the
four-gradient acting on the whole fermion bilinear.}
 \label{tb:discrete}
\end{table}
While the hermiticity condition does not impose any constraint, since one
can always multiply the individual bilinears by appropriate factors of $i$,
the ${\cal C}$ and ${\cal P}$ symmetries must be enforced.

Second, we observe that derivatives $\partial$ acting on the whole
bilinear are of order $Q$, while derivatives $\overleftrightarrow{\partial}$
acting inside a bilinear are of order $Q^0$ due to the presence
of the nucleon mass.  Therefore, at each order in the power
counting, only a finite number of $\partial$ appears, while it
is possible to have, in principle, any number of
$\overleftrightarrow{\partial}$. 
The situation is not so hopeless, however.  For instance,
the contracted product $\overleftrightarrow{\partial}_\mu \overleftrightarrow{\partial}^\mu$
inside a bilinear yields a squared mass term (without derivatives) plus
a $\partial_\mu \partial^\mu$ acting on the whole bilinear, which is
suppressed by $Q^2$.  Similarly, a term like
$\overleftrightarrow{\slashed{\partial}}\equiv \overleftrightarrow{\partial}_\mu\, \gamma^\mu$,
resulting from the contraction, in a bilinear, of $\overleftrightarrow{\partial}_\mu$
with one of the elements of the Clifford algebra, can be replaced by a term
without derivatives by making use of the equations of motion, {\it i.e.}
$i\, \slashed{\partial} \psi=m\, \psi$ and its adjoint; for example,
\begin{equation}
\label{eq:dsigma}
\overline{\psi}\, i\!\overleftrightarrow{\partial}_\mu \sigma^{\mu \nu}
\, \psi = 
 \, \overline{\psi}\,\left( \gamma^\nu \overrightarrow
{\slashed{\partial}} 
 + \overleftarrow {\slashed{\partial}} \gamma^\nu\right)\psi
 -\partial^\nu\left(\overline{\psi}\, \psi\right) 
= -\partial^\nu\left(\overline{\psi}\, \psi\right)\ .
\end{equation}
Hence, in general, no two Lorentz indices in a fermion bilinear can be
contracted with one another, except for the Levi-Civita tensor and
for the (suppressed) $\partial^2$ acting on the whole bilinear.

Some of the most problematic terms are of the type
\begin{equation}
  \label{eq:hn}
 \widetilde O^{(n)}_{\,\Gamma_A  \Gamma_B}=\frac{1}{(2m)^{2n}}
 (\overline{\psi}\, i\!\overleftrightarrow{\partial}^{\! \mu_1} \,
 i\!\overleftrightarrow{\partial}^{\! \mu_2} \cdots
 i\!\overleftrightarrow{\partial}^{\! \mu_n} \, \Gamma^\alpha_A \,\psi )\, 
   (\overline{\psi}\,
  i\!\overleftrightarrow{\partial}_{\!\mu_1} \,
  i\!\overleftrightarrow{\partial}_{\!\mu_2} 
  \cdots  i\!\overleftrightarrow{\partial}_{\!\mu_n} 
\, \Gamma_{B\, \alpha}\,   \psi )\ ,
\end{equation}
since, as stated above, $n$ can be any integer.  In fact,
a little thought shows that terms with $n>1$ do not introduce any new
operator structure up to ${\cal O}(Q^2)$ included.  This is
most easily seen by considering the matrix elements of such
terms between initial and final two-nucleon states with momenta,
respectively, ${\bf p}_1,{\bf p}_2$ and ${\bf p}_3,{\bf p}_4$.
These matrix elements consist of the product of two factors:
one given (in a schematic notation) by
$\left(\overline{u}_3\, \Gamma^\alpha_A\, u_1\right)
\, \left(\overline{u}_4\, \Gamma_{B\,\alpha}\, u_2\right)$---the
$u_i$ denote Dirac spinors---and another involving the
particles' four-momenta,
\begin{equation}
  \frac{[(p_1+p_3)\cdot(p_2+p_4)]^n}{(2m)^{2n}}\ ,
\end{equation}
which to ${\cal O}(Q^2)$ can be approximated as
\begin{equation}
  1+ \frac{n}{ 4m^2}
   \biggl[{\bf p}_1^2+{\bf p}_2^2+{\bf p}_3^2+{\bf p}_4^2-
   ({\bf p}_1+{\bf p}_3)\cdot({\bf p}_2+{\bf p}_4) \biggr]\ .
\end{equation}
Therefore, as discussed in more detail in the next section,
one only needs, in practice, to account for terms of type~(\ref{eq:hn})
with $n=0,1$.

\subsection{Lagrangian to order $Q^2$}
\label{sec:nr2}

Following the criteria laid out in the previous section, a complete
but non-minimal set consisting of 36 ${\cal P}$- and ${\cal C}$-conserving operators,
denoted as $\widetilde{O}_i$, is obtained.  They are listed in Table~\ref{tb:ops1}.
Note that some operator structures are missing, since they do not contribute at order $Q^2$.
For instance, operators having the $1 \otimes \gamma_5$ structure are at least of order $Q^4$:
${\cal P}$ symmetry requires the presence of an $\epsilon^{\mu\nu\alpha\beta}$ whose indices
(three of which space-like) must be contracted with partial derivatives, and an additional factor
$Q$ comes from the presence of $\gamma_5$, which mixes large and small components of
the Dirac spinors. 

\begin{table}[bth]
\begin{center}
\begin{tabular}{c|c|c}
\hline
\hline
$ 1 \otimes 1$ & $\,\,\,\widetilde{O}_1\,\,\,$&$(\overline{\psi} \psi) (\overline{\psi} \psi)$ \\
      & $\widetilde{O}_2$&$\frac{1}{4 m^2}(\overline{\psi} i\overleftrightarrow{\partial}^\mu
                                       \psi) (\overline{\psi}i\overleftrightarrow{\partial}_{\mu} \psi) $ \\
      & $\widetilde{O}_3$&$\frac{1}{4 m^2}(\overline{\psi} \psi )\partial^2 (\overline{\psi} \psi )$ \\
\hline
$\,\,1 \otimes \gamma\,\,\,\,\, \,$ & $\widetilde{O}_4$& $ \frac{1}{2 m} (\overline{\psi} 
                                            i\overleftrightarrow{\partial}^\mu \psi) (\overline{\psi}\gamma_\mu \psi) $ \\
      & $\widetilde{O}_5$& $\frac{1}{8 m^3} (\overline{\psi}i\overleftrightarrow{\partial}^\mu 
                                                i\overleftrightarrow{\partial}^\nu \psi )( \overline{\psi} \gamma_\mu 
                                                i\overleftrightarrow{\partial}_{\nu} \psi )$ \\
      & $\widetilde{O}_{6}$& $ \frac{1}{8m^3} (\overline{\psi}
                                            i \overleftrightarrow{\partial}_{\mu}  \psi)
                          \partial^2 (\overline{\psi} \gamma^\mu  \psi)  $ \\
\hline
$  1 \otimes \gamma \gamma_5 $ & $\widetilde{O}_7$& $ \frac{1}{8m^3} \epsilon_{\mu\nu\alpha\beta} 
                                              (\overline{\psi} i\overleftrightarrow{\partial}^\mu
                                                 \psi ) \partial^\nu
						 (\overline{\psi} i \overleftrightarrow{\partial}^\alpha
                                               \gamma^\beta \gamma_5 \psi) $ \\
\hline
$\gamma_5 \otimes \gamma_5 $ & $\widetilde{O}_8$& $(\overline{\psi} i\gamma_5 \psi) 
                                                     (\overline{\psi} i\gamma_5 \psi )$ \\
\hline
$ \gamma_5 \otimes \sigma$ & $\widetilde{O}_9$& 
        $ \frac{1}{4 m^2} \epsilon_{\mu \nu \alpha \beta} (\overline{\psi}  i\gamma_5 \psi)
        \partial^\mu( \overline{\psi} i\overleftrightarrow{\partial}^{\nu} \sigma^{\alpha \beta}\psi)$ \\
      & $\widetilde{O}_{10}$  & $ \frac{1}{4 m^2} \epsilon_{\mu \nu \alpha \beta}
	(\overline{\psi}  i\gamma_5 i\overleftrightarrow{\partial}^{\mu}
            \psi)
        \partial^\nu( \overline{\psi} \sigma^{\alpha \beta}\psi)$ \\
\hline
$\gamma \otimes \gamma $ & $\widetilde{O}_{11}$& $( \overline{\psi} \gamma^\mu \psi) ( \overline{\psi}
                                                     \gamma_\mu \psi ) $ \\
      & $\widetilde{O}_{12}$& $ \frac{1}{4 m^2} ( \overline{\psi} \gamma^\mu i\overleftrightarrow{\partial}^\nu \psi)
                                      (\overline{\psi}\gamma_\mu i\overleftrightarrow{\partial}_{\nu} \psi)$ \\   
      & $\widetilde{O}_{13}$& $\frac{1}{4 m^2}( \overline{\psi}    \gamma^\mu \psi) 
                                     \partial^2 (\overline{\psi} \gamma_\mu \psi) $ \\
      & $\widetilde{O}_{14}$& $\frac{1}{4 m^2} ( \overline{\psi}
				     \gamma^\mu
				     i\overleftrightarrow{\partial}^\nu \psi) 
                                                      (
				     \overline{\psi}\gamma_\nu
				     i\overleftrightarrow{\partial}_{\mu} \psi) $ \\ 
      & $\widetilde{O}_{15}$& $\frac{1}{16 m^4} ( \overline{\psi} \gamma^\mu i\overleftrightarrow{\partial}^\nu
                                            i\overleftrightarrow{\partial}^\alpha \psi) ( \overline{\psi} \gamma_\nu
                                            i\overleftrightarrow{\partial}_{\mu} 
                                            i\overleftrightarrow{\partial}_{\alpha} \psi) $ \\
\hline
$\gamma \otimes \gamma \gamma_5 $ & $\widetilde{O}_{16}$& $ 
                 \frac{1}{4 m^2} \epsilon_{\mu \nu \alpha \beta}( \overline{\psi}
                                           \gamma^\mu \psi)  \partial^\nu   (\overline{\psi} 
                                           i\overleftrightarrow{\partial}^\alpha
                                          \gamma^\beta \gamma_5 \psi)$ \\
      & $\widetilde{O}_{17}$& $  \frac{1}{4 m^2} \epsilon_{\mu \nu \alpha \beta}( \overline{\psi}
                                           \gamma^\mu i\overleftrightarrow{\partial}^\nu
                                           \psi)  \partial^\alpha   (\overline{\psi} 
                                          \gamma^\beta \gamma_5 \psi)$ \\
      & $\widetilde{O}_{18}$& $  \frac{1}{16 m^4} \epsilon_{\mu \nu \alpha \beta}( \overline{\psi}
                                           \gamma^\gamma i\overleftrightarrow{\partial}^\mu
                                           \psi)  \partial^\nu  
                                           (\overline{\psi} 
                                            i\overleftrightarrow{\partial}_\gamma
                                            i\overleftrightarrow{\partial}^\alpha
                                          \gamma^\beta \gamma_5 \psi)$ \\
\hline
$ \,\,\gamma \gamma_5 \otimes \gamma \gamma_5\,\,$ &
                                           $\widetilde{O}_{19}$& 
                                       $( \overline{\psi} \gamma^\mu \gamma_5 \psi)
                                       ( \overline{\psi} \gamma_\mu \gamma_5 \psi ) $ \\
      & $\widetilde{O}_{20}$& $ \frac{1}{4 m^2} ( \overline{\psi}
                                           \gamma^\mu \gamma_5  
                                           i \overleftrightarrow{\partial}^\nu 
                                          \psi) ( \overline{\psi}\gamma_\mu \gamma_5  
                                           i\overleftrightarrow{\partial}_\nu \psi)$ \\
     & $\,\,\,\widetilde{O}_{21}\,\,\,$& $ \frac{1}{4 m^2} ( \overline{\psi} \gamma^\mu \gamma_5  \psi)
                                                    \partial^2 (\overline{\psi} \gamma_\mu\gamma_5  \psi) $ \\
     & $\widetilde{O}_{22}$& $\frac{1}{4 m^2} ( \overline{\psi}  \gamma^\mu   \gamma_5  
                                              i \overleftrightarrow{\partial}^\nu
                                             \psi) ( \overline{\psi}\gamma_\nu\gamma_5 
                                              i\overleftrightarrow{\partial}_\mu \psi)$ \\
\hline
$ \,\,\gamma \gamma_5 \otimes \sigma\,\,\,\,\,$   & $\widetilde{O}_{23}$& $\frac{1}{4 m} \epsilon_{\mu\nu\alpha\beta}
                                                  ( \overline{\psi} \gamma^\mu \gamma_5 \psi)
                                            ( \overline{\psi} 
                                            i\overleftrightarrow{\partial}^\nu \sigma^{\alpha \beta} \psi )$ \\
     & $\widetilde{O}_{24}$& $ \frac{1}{16 m^3} \epsilon_{\mu\nu\alpha\beta} ( \overline{\psi} \gamma^\mu \gamma_5
                                               i\overleftrightarrow{\partial}^\gamma\psi)( \overline{\psi} 
                                               i\overleftrightarrow{\partial}^\nu  
                                               i\overleftrightarrow{\partial}_\gamma \sigma^{\alpha \beta} \psi)$ \\
     & $\widetilde{O}_{25}$& $\frac{1}{4 m} \epsilon_{\mu\nu\alpha\beta} ( \overline{\psi}\gamma^\mu \gamma_5
                                             i \overleftrightarrow{\partial}^\nu \psi) 
                                             ( \overline{\psi}  \sigma^{\alpha \beta} \psi) $ \\
     & $\widetilde{O}_{26}$& $\frac{1}{16 m^3} \epsilon_{\mu\nu\alpha\beta} ( \overline{\psi}\gamma^\mu \gamma_5
                                             i \overleftrightarrow{\partial}^\nu 
                                             i \overleftrightarrow{\partial}^\gamma \psi)
                                              ( \overline{\psi}
					     \sigma^{\alpha \beta}
                                             i\overleftrightarrow{\partial}_\gamma  \psi) $ \\
     & $\widetilde{O}_{27}$& $\frac{1}{16 m^3} \epsilon_{\mu\nu\alpha\beta}
                                                  ( \overline{\psi} \gamma^\mu \gamma_5 \psi)
                                            \partial^2( \overline{\psi} 
                                            i\overleftrightarrow{\partial}^\nu \sigma^{\alpha \beta} \psi )$ \\
     & $\widetilde{O}_{28}$& $\frac{1}{16 m^3} \epsilon_{\mu\nu\alpha\beta} ( \overline{\psi}\gamma^\mu \gamma_5
                                             i \overleftrightarrow{\partial}^\nu \psi) \partial^2
                                             ( \overline{\psi}  \sigma^{\alpha \beta} \psi) $ \\
     & $\widetilde{O}_{29}$& $\frac{1}{16 m^3} \epsilon_{\mu\nu\alpha\beta}
                                             (\overline{\psi}\gamma^\gamma \gamma_5 
                                             i \overleftrightarrow{\partial}^\mu
                                             \psi) (
                                             \overline{\psi} 
                                            i\overleftrightarrow{\partial}_\gamma
                                            i\overleftrightarrow{\partial}^\nu 
                                            \sigma^{\alpha \beta} \psi )$ \\
     & $\widetilde{O}_{30}$& $\frac{1}{16 m^3} \epsilon_{\mu\nu\alpha\beta}
                                             (\overline{\psi}\gamma^\mu \gamma_5 
                                             i \overleftrightarrow{\partial}^\nu
                                             i \overleftrightarrow{\partial}_\gamma
                                             \psi) (
                                             \overline{\psi} 
                                            i\overleftrightarrow{\partial}^\alpha
                                            \sigma^{\beta\gamma} \psi )$ \\
$\sigma \otimes \sigma $   & $\widetilde{O}_{31}$& $ \frac{1}{2}( \overline{\psi} \sigma^{\mu\nu} \psi)
                                                ( \overline{\psi} \sigma_{\mu\nu}\psi)$ \\
     & $\widetilde{O}_{32}$& $\frac{1}{8 m^2} ( \overline{\psi} \sigma^{\mu\nu}i\overleftrightarrow{\partial}^\alpha
                                            \psi) (\overline{\psi}\sigma_{\mu\nu}
                                             i\overleftrightarrow{\partial}_\alpha
                                             \psi) $ \\ 
     & $\widetilde{O}_{33}$& $ \frac{1}{8 m^2}( \overline{\psi} \sigma^{\mu\nu} \psi )\partial^2 (\overline{\psi}
                                            \sigma_{\mu\nu}\psi)$ \\
     & $\widetilde{O}_{34}$& $ \frac{1}{8 m^2} ( \overline{\psi}
                                             \sigma^{\mu \alpha} 
                                          i\overleftrightarrow{\partial}^\nu
                                          \psi) ( \overline{\psi} 
                                          i\overleftrightarrow{\partial}_\alpha
                                          \sigma_{\mu \nu} \psi)$ \\
     & $\widetilde{O}_{35}$& $ \frac{1}{32m^4} \epsilon_{\mu\nu \gamma\delta} \epsilon_{\alpha \beta \rho \sigma}
                                            ( \overline{\psi} \sigma^{\mu\nu} i\overleftrightarrow{\partial}^\gamma
                                             i\overleftrightarrow{\partial}^\rho \psi) ( \overline{\psi}
                                             \sigma^{\alpha \beta}
                                              i\overleftrightarrow{\partial}^\delta i\overleftrightarrow{\partial}^\sigma 
                                             \psi)$ \\
     & $\widetilde{O}_{36}$& $ \frac{1}{32m^4} \epsilon_{\mu\nu \gamma\delta} \epsilon_{\alpha \beta \rho \sigma}
                                            ( \overline{\psi} \sigma^{\mu\nu} i\overleftrightarrow{\partial}^\gamma
                                             i\overleftrightarrow{\partial}^\rho \psi) 
                                             \partial^\delta\partial^\sigma( \overline{\psi}
                                             \sigma^{\alpha \beta}
                                             \psi)$ \\
\hline
\hline
\end{tabular}
\caption{A complete, but non-minimal, set of  relativistic contact interactions.  Note that
the field operator products are understood to be normal-ordered.}
\label{tb:ops1}
\end{center}
\end{table}
The non-relativistic reduction of the $\widetilde{O}_i$ up
to terms of order $Q^2$ included is tedious but straightforward.
To this end, the relativistic field (specifically, its positive-energy component, where
the (+) superscript has been dropped for simplicity)
\begin{equation}
  \psi(x)= \int \frac{d {\bf p}}{(2 \pi)^3}
   \frac{m}{E_p}\, \widetilde{b}_{s}({\bf p}) \,u^{(s)}({\bf p})\,
  {\mathrm{e}}^{-i p\cdot  x} ,
\end{equation}
with normalizations
\begin{equation}
  \left[ \widetilde{b}_s({\bf p})\, ,\, \widetilde{b}_{s^\prime}^{\, \dagger} ({\bf p}^\prime) \right]_+= 
  \frac{E_p}{m}\, (2\pi)^3\delta({\bf p}-{\bf p}^\prime)\,
  \delta_{ss^\prime} \ , \quad \overline{u}^{\,(s)} ({\bf p})
  u^{(s')}({\bf p}) = \delta_{s s'} \ ,
\end{equation}
is expanded in terms of the non-relativistic field $N(x)$, defined in Eq.~(\ref{eq:nrf}), as
\begin{equation}
  \psi(x) = \left[ \left( \begin{array}{c} 1 \\
                                           0 \end{array} \right) -
   \frac{i}{2m}\left( \begin{array}{c} 0 \\
  {\bm \sigma} \cdot {\bm \nabla} \end{array} \right)
  +\frac{1}{8m^2}
   \left( \begin{array}{c} {\bm \nabla}^2\\
                              0 \end{array} \right)\right] N(x) + {\cal O}(Q^3)  \ .\label{eq:psiN}
\end{equation}
Note that the relativistic ($\widetilde{b}$) and non-relativistic ($b$)
versions of the annihilation operator are
related to each other by $b_s ({\bf p}) = \sqrt{m/E_p}\, \,\widetilde{b}_s ({\bf p})$.
Partial integrations and use of the fields' equations of motion to eliminate time
derivatives,
\begin{eqnarray}
 \frac{i}{2m}\overline{\psi}\, \overleftrightarrow{\partial}^0 \,\psi&=&  \psi^\dagger \left[
1-\frac{i}{2m} {\bm \gamma}\cdot \left(\overleftarrow{\bm \nabla}+
 \overrightarrow{\bm \nabla}\right)\right]\psi \nonumber \\
&=&N^\dagger N-\frac{1}{8m^2}N^\dagger\left[\left( \overleftarrow{\bm \nabla}+
\overrightarrow{\bm \nabla}\right)^2
+2\, i\, {\bm \sigma}\cdot\overleftarrow{\bm \nabla}\times \overrightarrow{\bm \nabla} \right]N
+ {\cal O}(Q^3) \ , \label{eq:timed}
\end{eqnarray}
lead to the operators $\widetilde{O}_i^{\rm NR}$ of Table~\ref{tb:ops2}.  They are given there as linear
combinations of the operator basis $O_i$, defined previously
(see Table~\ref{tb:tab1}).

Returning briefly to the discussion of the terms of type
$\widetilde{O}^{(n)}_{\,\Gamma_A \Gamma_B}$ in Eq.~(\ref{eq:hn}),
it is useful to separate such terms into three classes,
depending on whether the non-relativistic expansion of the
spinor matrix element
$(\overline{u}_3\, \Gamma^\alpha_A u_1)\,
(\overline{u}_4\, \Gamma_{B\, \alpha} u_2)$ is i) $1+{\cal O}(Q^2)$
(class I), or ii) $\pm {\bm \sigma}_1\cdot {\bm \sigma}_2+{\cal O}(Q^2)$
(class II), or iii) ${\cal O}(Q^2)$ (class III).
Making use of the relations~(\ref{eq:psiN}) and
(\ref{eq:timed}), we find that terms in class I
 reduce to
\begin{equation}
\widetilde{O}^{(n)}_{\,\Gamma_A \Gamma_B}=\widetilde{O}^{(n=0)}_{\,
  \Gamma_A \Gamma_B}
 + \frac{n}{4m^2} ( O_1 - 2\, O_2 - 2 \, O_3) +{\cal O}(Q^4)\ ,
\label{eq:hn1}
\end{equation}
while those in class II  reduce to
\begin{equation}
\widetilde{O}^{(n)}_{\,\Gamma_A \Gamma_B}=\widetilde{O}^{(n=0)}_{\,
  \Gamma_A \Gamma_B}
 \pm \frac{n}{4m^2} \left(3\, O_9 +2\, O_{12} +2 \, O_{14}\right)+{\cal O}(Q^4)\ ,
\label{eq:hn1a}
\end{equation}
and lastly the terms in class III  are
simply given, for any $n$, by $\widetilde{O}^{(n=0)}_{\, \Gamma_A \Gamma_B}$
up to corrections ${\cal O}(Q^4)$.  A quick glance
at Tables~\ref{tb:ops1} and~\ref{tb:ops2} shows
that the relations above are verified: consider, for example,
$\widetilde{O}_1$ and $\widetilde{O}_2$ in class I,
$\widetilde{O}_{20}$ and $\widetilde{O}_{21}$ in class II,
and $\widetilde{O}_8$ in class III, and their corresponding non-relativistic
reductions.

\begin{table}[bth]
\begin{center}
\begin{tabular}{c|c}
\hline
\hline
$\,\,\,\widetilde{O}^{\rm NR}_1\,\,\,\,\,$ & $ O_S+\frac{1}{4 m^2} \left( O_1 + 2\, O_2 + 2\, O_3 + 2\, O_5 \right) $ \\
$\,\,\,\widetilde{O}^{\rm NR}_2\,\,\,$ & $O_S +\frac{1}{4 m^2} \left( 2\, O_1 + 2\, O_5 \right) $ \\
$\,\,\,\widetilde{O}^{\rm NR}_3\,\,\,$ & $\frac{1}{4 m^2} \left(O_1
+2\, O_2\right)  $ \\
\hline
$\,\,\,\widetilde{O}^{\rm NR}_4\,\,\,$ & $ O_S+\frac{1}{4 m^2} \left( O_1 -2\, O_2 + 2\,O_6 \right) $ \\
$\,\,\,\widetilde{O}^{\rm NR}_5\,\,\,$ & $ O_S+\frac{1}{4 m^2} \left(
2\, O_1 -4\, O_2 -2\, O_3 + 2\,O_6 \right) $ \\
$\,\,\,\widetilde{O}^{\rm NR}_6\,\,\,$ & $ \frac{1}{4 m^2} \left( O_1+
2\, O_2\right)  $ \\
\hline
$\,\,\,\widetilde{O}^{\rm NR}_7\,\,\,$ & $ \frac{1}{4
  m^2}\left(-2\,O_5+2\,O_6\right) $ \\
\hline
$\,\,\,\widetilde{O}^{\rm NR}_8\,\,\,$ & $ \frac{1}{4 m^2}\left(O_7
+2\, O_{10}\right)   $ \\
\hline
$\,\,\,\widetilde{O}^{\rm NR}_9\,\,\,$ & $ \frac{1}{4 m^2}\left(2\, O_7 +4\, O_{10} \right)$ \\
$\,\,\,\widetilde{O}^{\rm NR}_{10}\,\,\,$ & $ \frac{1}{4 m^2}\left(
-2\, O_7 - 4\, O_{10} \right)$ \\
\hline
$\,\,\,\widetilde{O}^{\rm NR}_{11}\,\,\,$ & $ O_S +\frac{1}{4 m^2}
  \left(-4\, O_2 -2\, O_5 +4\, O_6 +O_7 - O_9 + 2\, O_{10} - 2\, O_{12} \right)$ \\
$\,\,\,\widetilde{O}^{\rm NR}_{12}\,\,\,$ & $ O_S +\frac{1}{4 m^2}
  \left( O_1 -6\, O_2 - 2\, O_3 -2\, O_5 + 4\,  O_6 + O_7 -  O_9 +2\, O_{10} -2\, O_{12} \right)$ \\
$\,\,\,\widetilde{O}^{\rm NR}_{13}\,\,\,$ & $  \frac{1}{4 m^2}\left(O_1 + 2\, O_2\right) $ \\
$\,\,\,\widetilde{O}^{\rm NR}_{14}\,\,\,$ & $ O_S +\frac{1}{4 m^2}
  \left( O_1 -6 \,O_2 - 2 \,O_3 -2\, O_5 + 4 \, O_6  \right)$ \\ 
$\,\,\,\widetilde{O}^{\rm NR}_{15}\,\,\,$ & $ O_S +\frac{1}{4 m^2}
  \left(2\, O_1 -8\, O_2 -4\, O_3 - 2\, O_5 +4\,  O_6  \right)$ \\ 
\hline
$\,\,\,\widetilde{O}^{\rm NR}_{16}\,\,\,$ & $\frac{1}{4 m^2}
\left(-2\,O_5 + 2\, O_6 + O_7 - O_9 + 2\, O_{10} - 2\, O_{12}\right) $ \\ 
$\,\,\,\widetilde{O}^{\rm NR}_{17}\,\,\,$ & $\frac{1}{4 m^2}
\left(-O_7 +O_9  - 2\, O_{10} + 2\, O_{12}\right) $ \\
$\,\,\,\widetilde{O}^{\rm NR}_{18}\,\,\,$ & $\frac{1}{4 m^2}
\left( -2\,O_5 + 2\, O_6 \right) $ \\
\hline
$\,\,\,\widetilde{O}^{\rm NR}_{19}\,\,\,$ & $ -O_T -\frac{1}{4 m^2} \left(- 2 \, O_6 + O_7 - O_9 - 2 \, O_{10} - 2\, O_{12} + 2\,
                                                                               O_{13} - 2\, O_{14} \right)$ \\
$\,\,\,\widetilde{O}^{\rm NR}_{20}\,\,\,$ & $-O_T -\frac{1}{4 m^2} \left( -2\, O_6 + O_7 +2\,  O_9 - 2\, O_{10}+      2\, O_{13} \right)  $ \\
$\,\,\,\widetilde{O}^{\rm NR}_{21}\,\,\,\,\,$ & $\frac{1}{4 m^2} \left( -O_9 - 2 \, O_{12} \right)$ \\
$\widetilde{O}^{\rm NR}_{22}$ & $  \frac{1}{4 m^2} \left( -2\, O_7
-2\, O_8 - 4\, O_{13} \right)$ \\
\hline
$\widetilde{O}^{\rm NR}_{23}$ & $  -O_T  - \frac{1}{4m^2} \left( -2\, O_6 +2\, O_7  - O_9  -2\, O_{12} +2\, O_{13} -2\, O_{14} \right)$ \\
$\widetilde{O}^{\rm NR}_{24}$ & $-O_T -\frac{1}{4 m^2} \left( -2\, O_6 +2\, O_7  + 2\, O_9  +2\, O_{13} \right) $ \\
$\widetilde{O}^{\rm NR}_{25}$ & $-O_T -\frac{1}{4 m^2} \left(  - 2\,
O_5 -2\, O_8 + O_9  - 2 \, O_{12} - 2 \, O_{13} \right) $ \\
$\widetilde{O}^{\rm NR}_{26}$ & $ -O_T -\frac{1}{4m^2} \left(   - 2\, O_5  -2\, O_8 +4 \, O_9 -2\, O_{13} +2\, O_{14} \right)$ \\
$\widetilde{O}^{\rm NR}_{27}$ & $\frac{1}{4m^2} \left( - O_9 -2 \, O_{12}\right)$ \\
$\widetilde{O}^{\rm NR}_{28}$ & $\frac{1}{4m^2} \left( - O_9 -2 \, O_{12}\right)$ \\
$\widetilde{O}^{\rm NR}_{29}$ & $\frac{1}{4m^2} \left(  -2 \, O_7 -
2\, O_8 - 4\, O_{13} \right)$ \\
$\widetilde{O}^{\rm NR}_{30}$ & $\frac{1}{4 m^2} \left(   -O_5 +  O_6
-  O_7 - O_8 +2\, O_9 -2\, O_{13} + 2\, O_{14}\right) $ \\
$\widetilde{O}^{\rm NR}_{31}$ & $ O_T+ \frac{1}{4 m^2} \left( -O_1 -
2\, O_2  - 4\, O_5 + 2\, O_6  + O_7 - 2\, O_8 +2\, O_{10} -4\, O_{12}
- 2\, O_{13} \right)$ \\ 
$\widetilde{O}^{\rm NR}_{32}$ & $ O_T + \frac{1}{4 m^2} \left( -O_1 -
2\, O_2  - 4\, O_5  + 2\, O_6 +O_7 - 2 \, O_8  +3\, O_9 +2\, O_{10} -2\, O_{12} - 2\, O_{13} + 2\, O_{14} \right)$ \\
$\widetilde{O}^{\rm NR}_{33}$ & $ \frac{1}{4 m^2} \left(  O_9 + 2 \,
O_{12} \right) $ \\
$\widetilde{O}^{\rm NR}_{34}$ & $ \frac{1}{4 m^2} \left(  -\frac{1}{2}
O_1 -  O_2 - 2\, O_5 + 2\, O_6 - O_7 -  O_8 +2\, O_9 - 2\, O_{13} +
2\, O_{14} \right) $ \\
$\widetilde{O}^{\rm NR}_{35}$ & $\frac{1}{4 m^2} \left(  4\, O_7 + 4\, O_8 + 8\, O_{13}  \right) $ \\
$\widetilde{O}^{\rm NR}_{36}$ & $\frac{1}{4 m^2} \left(  2\, O_7 +  4\, O_{10}  \right) $ \\
\hline
\hline
\end{tabular}
\caption{The non-relativistic expressions, up to order $Q^2$ included,
corresponding to the contact interactions of Table~\protect\ref{tb:ops1}.}
\label{tb:ops2}
\end{center}
\end{table}
Inspection of Table~\ref{tb:ops2} shows that a set of linearly independent
operator combinations can be defined as
\begin{equation} 
\label{eq:7op}
\begin{array}{l} 
O_S + (O_1 + O_3 + O_5 + O_6)/(4m^2)\\
O_T -( O_5 +  O_6 - O_7 + O_8 +  2\, O_{12} +  O_{14})/(4m^2) \\
O_1 + 2\, O_2\\
2\, O_2 + O_3\\
O_9 + 2\, O_{12}\\
O_9 + O_{14}\\
 O_5 - O_6\\
O_7 + 2\, O_{10}\\
O_7 + O_8  + 2\, O_{13}\\
\end{array}
\end{equation}
consisting of 2 leading (of order $Q^0$) and 7 sub-leading ($Q^2$) ones,
in agreement with the results of an analysis based on the heavy-baryon
formulation of EFT~\cite{Epelbaum00,Girlanda09}, so that the effective
Lagrangian can be  written as
\begin{eqnarray}
{\cal L} &=& -\frac{1}{2} C_S \Bigg[  O_S + \frac{1}{4 m^2} ( O_1 + O_3 +
  O_5 + O_6 ) \Bigg] - \frac{1}{2} C_T \Bigg[ O_T - \frac{1}{4 m^2} 
  \Bigg(O_5  + O_6  -  O_7  + O_8   \nonumber \\
&& + 2\, O_{12}+ O_{14} \Bigg) \Bigg] 
  - \frac{1}{2} C_1 ( O_1 + 2\, O_2) + \frac{1}{8} C_2 (2\, O_2
+ O_3 ) - \frac{1}{2} C_3 ( O_9 + 2\, O_{12})  \nonumber \\
&&- \frac{1}{8} C_4 (O_9 +
O_{14})+ \frac{1}{4} C_5 ( O_6 - O_5) - \frac{1}{2} C_6 (O_7 + 2 \,
O_{10}) -\frac{1}{16} C_7 (O_7 + O_8 + 2\, O_{13})\, . \nonumber \\
&&
\end{eqnarray}
Evaluation of the
matrix elements of the operators $O_i$ between initial and final two-nucleon
states with momenta ${\bf P}/2+{\bf p},{\bf P}/2-{\bf p}$ and 
${\bf P}/2+{\bf p}^\prime,{\bf P}/2-{\bf p}^\prime$, {\it i.e.}
\begin{equation}
O_i({\bf p}^\prime, {\bf p};{\bf P}) =\int d{\bf x}\, 
\langle {\bf P}/2+{\bf p}^\prime,{\bf P}/2-{\bf p}^\prime\mid
O_i \mid {\bf P}/2+{\bf p},{\bf P}/2-{\bf p}\rangle \ ,
\end{equation}
shows that the 7 sub-leading combinations above give vanishing ${\bf P}$-dependent
contributions, and in fact lead, in the center-of-mass frame, to the 7
${\bf k}$- and ${\bf K}$-dependent terms occurring in $v^{\rm CT2}({\bf k}, {\bf K})$,
Eq.~(\ref{eq:ct2}).  Similarly, the 2 leading combinations and associated $1/m^2$
corrections---first 2 lines of Eq.~(\ref{eq:7op})---give rise, respectively, to the ${\bf P}$-dependent terms
\begin{equation}
-\frac{P^2}{2m^2}+\frac{i}{4m^2}\left({\bm \sigma}_1-{\bm \sigma}_2\right)
\cdot {\bf P}\times{\bf k}
\end{equation}
and
\begin{equation}
-\frac{P^2}{2m^2}\, {\bm \sigma}_1\cdot{\bm \sigma}_2
-\frac{i}{4m^2}\left({\bm \sigma}_1-{\bm \sigma}_2\right)
\cdot {\bf P}\times{\bf k}
+\frac{1}{m^2}\left({\bm \sigma}_1\cdot{\bf P}\,\,{\bm \sigma}_2\cdot {\bf K}
-{\bm \sigma}_1\cdot{\bf K}\,\,{\bm \sigma}_2\cdot {\bf P}\right) \ ,
\end{equation}
which, after multiplication by $C_S/2$ and $C_T/2$, are precisely the terms
entering the potential $v^{\rm CT2}_{\bf P}({\bf k}, {\bf K})$ in Eq.~(\ref{eq:ct2p}).
\section{Poincar\'e algebra constraints}

As an alternative to the procedure discussed in the previous section, one can
impose the Poincar\'e algebra constraints on the Hamiltonians derived
from the Lagrangians in Eqs.~(\ref{eq:op0lagr}) and~(\ref{eq:op2lagr}),
${\cal H}^{(n)}_I = - {\cal  L}^{(n)}_I$.  In the instant
form of relativistic dynamics, the interactions affect not only the Hamiltonian
$H$ but also the boost generators ${\bf K}$.  We write
\begin{equation}
H=H_0 + H_I, \quad {\bf K} = {\bf K}_0 + {\bf K}_I, \quad {\bf P} =
{\bf P}_0, \quad {\bf J}= {\bf J}_0,
\end{equation}
to distinguish between the operators in the absence (with subscript 0) and in the
presence (without subscript) of interactions, and impose the following commutation
relations:
\begin{equation}
\label{eq:poincare}
\begin{array}{llll}
\left[ J^i\, , \, J^j \right] = i\, \epsilon^{ijk} J^k\ , & \left[K^i\, ,\, K^j\right] = - i\,
\epsilon^{ijk} J^k\ , & \left[J^i\, ,\, K^j\right] = i\, \epsilon^{ijk} K^k \ , 
& \left[P^\mu\, , \, P^\nu\right]=0 \ ,\\
 \left[K^{i}\, , \, P^j\right] =  i\, \delta^{ij} H\ , & \left[J^i \, , \, P^j\right] = i \, \epsilon^{ijk}
P^k\ , & {\left[K^i\, ,\, H\right] =  i\, P^i}, &  \left[J^i \, ,\, H \right]=0\ .
\end{array}
\end{equation}
The free Lorentz boost generators are derived from the  energy-momentum
tensor of the free fermionic theory
$T^{\mu\nu}=(i/2)\, \overline\psi\,  \gamma^\mu\,
\overleftrightarrow{\partial}^\nu \psi$ as
\begin{equation}
K_0^i= \int d{\bf x}\, \left(  x^i \, T^{\,00} - t\, T^{\,0i\,}  \right) \ ,
\end{equation}
where, for the time being, $\psi$ denotes the field with both positive-
and negative-energy components.  The use of the symmetric
energy-momentum tensor, the Belinfante tensor~\cite{weinbergbook},
\begin{equation}
\Theta^{\mu\nu}=T^{\mu\nu}   +\frac{1}{8} \partial_\alpha\, \overline{\psi}
\left[ \gamma^\alpha\, , \, \sigma^{\mu\nu} \right]_+ \psi
\end{equation}
would make no difference.  Insertion of the field expansions
in terms of normal modes in the equation above and manipulations of the
resulting expressions lead to
\begin{eqnarray}
\label{eq:freeboost}
K_0^i=\frac{i }{2} \int \frac{d{\bf p}}{(2 \pi)^3}\frac{m}{E_p} 
\Bigg[ E_p\left[ \widetilde{b}^\dagger_s ({\bf p})  \overleftrightarrow{\nabla}^i_{\!\! \bf p} \,
\widetilde{b}_s ({\bf p}) \right] \!\!
&+&\!\!m\, \widetilde{b}^\dagger_s ({\bf p})\, \widetilde{b}_{s'} ({\bf p}) 
\left[ u^{(s) \dagger } ({\bf p}) \overleftrightarrow{\nabla}^i_{\!\!\bf p}\, u^{(s')} ({\bf p})\right] \nonumber \\
+\,E_p \left[\widetilde{d}^{\, \dagger}_s ({\bf p})  \overleftrightarrow{\nabla}_{\!\!\bf p}^i \, 
\widetilde{d}_s ({\bf p})\right]\!\!
&-&\!\!m\,  \widetilde{d}^{\,\dagger}_{s'} ({\bf p})\, \widetilde{d}_s ({\bf p})
\left[ v^{(s)\dagger } ({\bf p}) \overleftrightarrow{\nabla}^i_{\!\! \bf p} \, v^{(s')} ({\bf p}) \right] \Bigg]\ ,
\end{eqnarray}
where $\widetilde{d}$ and $\widetilde{d}^{\, \dagger}$ are
annihilation and creation operators for antinucleons, and
$\nabla^i_{\bf p}$ denotes a derivative with respect to $p^i$.
Note that $K^i_0$ is time independent, since it is the spatial integral
of the time component ($\rho=0$) of a conserved current, $\partial_\rho M^{\rho\, 0\, i} = 0$
with $M^{\rho \mu \nu} = x^\nu \Theta^{\rho \mu} - x^\mu \Theta^{\rho \nu}$.
By making use of
\begin{equation}
\left[ {\bf K}_0\, , \, \widetilde{b}_s ({\bf p}) \right] = - i\, E_p
{\bm \nabla}_{\bf p}\, \widetilde{b}_s ({\bf p}) - \frac{1}{2\, ( m + E_p) }\, {\bf p} \times {\bm \sigma}_{s s'} \, \widetilde{b}_{s'} ( {\bf p}) \ ,
\label{eq:bk0}
\end{equation}
and a similar relation for $\widetilde{d}_s({\bf p})$, in which the only difference is
the sign of the second term on the right-hand-side of Eq.~(\ref{eq:bk0}), one can show
that the following commutation relations between the ``free'' generators are fulfilled
\begin{equation}
\left[K_0^{i}\, , \, P^j\right] =  i\, \delta^{ij} H_0 \ , \qquad \left[{\bf K}_0\, ,\, H_0\right] =  i\, {\bf P} \ ,
\end{equation}
where
\begin{equation}
\left(\!\!\begin{array}{c} {\bf P} \\
H_0 \end{array}\!\! \right)
= \int \frac{d{\bf p}}{(2 \pi)^3}
\frac{m}{E_p}  \left(\!\! \begin{array}{c} {\bf p} \\
E_p \end{array}\!\! \right) \left[ \widetilde{b}_s^\dagger ({\bf p})\, \widetilde{b}_s ({\bf p}) +
\widetilde{d}_s^{\,\dagger} ({\bf p}) \widetilde{d}_s ({\bf p}) \right] \ .
\end{equation}

We now turn our attention to the interacting theory.
The addition of an interaction term
\begin{equation}
H_I=\int d{\bf x}\, {\cal H}_I (t=0,{\bf x})\ ,
\end{equation}
requires the addition of a corresponding term ${\bf K}_I$ in the boost generators, as inspection of
the first commutator on the second line of Eq.~(\ref{eq:poincare}) makes clear.  Quite generally,
this term can be expressed as ${\bf K}_I = {\bf W} + \delta {\bf W}$,
where 
\begin{equation}
{\bf W} = \int d{\bf x} \, \,{\bf x}\, {\cal H}_I (0,{\bf x}) \ ,
\end{equation}
and $\delta {\bf W}$ is translationally invariant, {\it i.e.}~$\left[\delta W^i\, , \,P^j\right]=0$.
This latter condition ensures that the  commutator
$\left[K^i\, ,\, P^j\right]=i \, \delta^{ij}\, H$ is satisfied, since
$\left[W^i\, , \,P^j\right]=i \,\delta^{ij}\, H_I$.  A ``minimal''
choice would correspond to the case $\delta\, {\bf W} = 0$~\cite{Krajcik74,Carlson93}.

In order to proceed systematically, it is useful to introduce the following low-energy
power counting
\begin{equation}
 H_0 \sim Q^0+{\cal O}(Q^2)\ ,
 \quad {\bf P} \sim Q\ , \quad {\bf J} \sim Q^0\ , 
\quad {\bf K}_0 \sim Q^{-1}+{\cal O}(Q^1)\ ,
\label{eq:hcnt}
\end{equation}
which follows by observing that $\widetilde{b}$ and $\widetilde{b}^\dagger$, as well
as their non-relativistic counterparts $b$ and $b^\dagger$, each scale
as $Q^{-3/2}$, and by expanding $E_p$ and the Dirac spinors
in powers of $p/m$.  We now
require that the commutation relations among the Poincar\'e group generators be
satisfied order by order in this power counting.
To this end, it is useful to express
\begin{equation}
\label{eq:ek0h0}
{\bf K}_0={\bf K}_0^{(-1)}+{\bf K}_0^{(1)} +\dots \ ,\qquad
H_0=H_0^{(0)}+H_0^{(2)} +\dots \ ,
\end{equation}
where the superscript $(n)$ denotes the order in our power
counting, that is ${\bf K}_0^{(n)}, H_0^{(n)} \sim Q^n$.
Then the commutators of ${\bf K}_0^{(n)}$ and $H_0^{(n)}$
with the non-relativistic annihilation operator $b_s({\bf p})$
read at leading order as
\begin{equation} \label{eq:lcomm}
 \left[ {\bf K}^{(-1)}_0 \, , \, b_s ({\bf p} )\right] = -i \,m\,
{\bm \nabla}_{\bf p} \, b_s ({\bf p}) \ , \qquad
\left[ H^{(0)}_0 \, , \, b_s ({\bf p} )\right] = -m\, b_s ({\bf p})\ ,
\end{equation}
and at next to leading order as
\begin{eqnarray} 
\label{eq:k0b}
\left[ {\bf K}^{(1)}_0 \, , \, b_s ({\bf p} )\right] &=& -i\, 
\frac{p^2}{2 m} \,{\bm \nabla}_{\bf p} \, b_s ({\bf p}) 
- \frac{1}{4 m} {\bf p} \times {\bm \sigma}_{s s'} \, b_{s'}({\bf p})
-i\, \frac{{\bf p}}{2m}b_s({\bf p})\ ,\\
\left[ H^{(2)}_0 \, , \, b_s ({\bf p})\right] &=& -\frac{p^2}{2m}
\, b_s ({\bf p})\ ,
\end{eqnarray}
where the last term in Eq.~(\ref{eq:k0b})
comes from the gradient ${\bm \nabla}_{\bf p}$
acting on the factor $\sqrt{E_p/m}$
relating $\widetilde{b}_s({\bf p})$ to $b_s({\bf p})$.
It can now be shown that only the sub-leading terms
of $\left[ K^i_0 \, , \, K_0^j \right]$
and $\left[ {\bf K}_0\, , \, H_0\right] $,
respectively of order $Q^0$ and $Q^1$, are non-vanishing,
consistently with the power counting for
the angular momentum (${\bf J}$) and linear momentum (${\bf P}$) operators, established in Eq.~(\ref{eq:hcnt}).

We write the interaction Hamiltonian as
\begin{equation}
H_I=H^{(3)} + H^{(5)} \ ,
\end{equation}
where $H^{(3)}$ and $H^{(5)}$ are obtained from the Lagrangians in Eqs.~(\ref{eq:op0lagr}) and~(\ref{eq:op2lagr}), and the superscripts
denote the order in our power counting.  Correspondingly, we have
\begin{equation}
{\bf W} ={\bf W}^{(2)} + {\bf W}^{(4)}\ .
\end{equation}
Assuming, for the time being, $\delta {\bf W}=0$, the relations to satisfy are
\begin{eqnarray}
\!\!\!\!\left[ K_0^i + W^{(2)i} + W^{(4)i} +\dots\, , \, K_0^j + W^{(2)j} +
  W^{(4)j} + \dots \right]\!\! &=&\!\! -i\, \epsilon_{ijk}\, J^k = 
\left[ K_0^i \, , \, K_0^j \right] \ , \label{eq:kkrel}\\
\!\!\!\!
\left[ {\bf K}_0 + {\bf W}^{(2)} + {\bf W}^{(4)} + \dots\, , \, H_0 +
  H^{(3)} + H^{(5)} + \dots \right]\!\! &=& \!\! i\, {\bf P} = 
\left[ {\bf K}_0 \, , \, H_0\right] \label{eq:khrel}\ ,
\end{eqnarray}
where the $\dots$ represent additional terms to be determined below.
These relations impose non trivial constraints on $H^{(3)}$ and $H^{(5)}$.
We first examine those on $H^{(3)}$.

By expanding ${\bf K}_0$ and $H_0$
as in Eq.~(\ref{eq:ek0h0}), we find that the leading order relations
\begin{eqnarray}
 \label{eq:comm43}
\left[ K_0^{(-1)i}\, , \, W^{(2)j} \right] + \left[W^{(2)i}\, ,\, K_0^{(-1)j} \right]&=&0 \ , \\
\label{eq:comm43a}
\left[ {\bf K}^{(-1)}_0\, , \, H^{(3)} \right] 
+\left[ {\bf W}^{(2)}\, , \, H^{(0)}_0 \right] &=&0 \ ,
\end{eqnarray}
are fulfilled (see Appendix~\ref{app:a1}), so that inclusion
of the (leading) contact Hamiltonian
\begin{equation}
\label{eq:leadingH}
H^{(3)} = \frac{1}{2}\int d{\bf x} \, \left (C_S\, O_S+C_T\, O_T\right)   \equiv 
C_S \, H_S^{(3)} + C_T\, H_T^{(3)}
\end{equation}
does not spoil the Poincar\'e covariance of the theory (in leading order)---the
operators $O_S$, $O_T$, and $O_i$ are those defined in Table~\ref{tb:tab1} .  At next-to-leading
order, we may split $H^{(5)}$ as 
\begin{equation}
H^{(5)}=H_1^{(5)} + H_2^{(5)},
\end{equation}
where $H_1^{(5)}$ and the corresponding ${\bf W}_1^{(4)}$ are found by
imposing the relations
\begin{eqnarray}
 \label{eq:comm44}
\left[ K_0^{(1)i}\, , \, W^{(2)j} \right] + \left[W^{(2)i}\, ,\, K_0^{(1)j} \right] +
  \left[K_0^{(-1)i}\, ,\, W_1^{(4)j}\right] + 
\left[W_1^{(4)i}\, , \, K_0^{(-1)j} \right] &=&0\ , \\
\label{eq:comm44a}
\left[ {\bf K}^{(1)}_0\, , \, H^{(3)} \right] +
\left[ {\bf K}^{(-1)}_0\, , \, H_1^{(5)}  \right] + 
\left[ {\bf W}^{(2)}\, , \, H^{(2)}_0 \right] + 
\left[ {\bf W}_1^{(4)}\, , \, H^{(0)}_0\right] &=&0 \ .
\end{eqnarray}
 After
some algebra (see Appendix~\ref{app:a1}), we find that
\begin{equation}
\label{eq:driftH}
H^{(5)}_{1}=C_S \,H^{(5)}_S+ C_T\, H^{(5)}_T \ ,
\end{equation}
with
\begin{eqnarray}
\label{eq:h5s}
H^{(5)}_S&=&\frac{1}{8 m^2} \int\! d{\bf x}\, \left( O_1 + O_3 + O_5 +
O_6 \right) \ , \\
\label{eq:h5t}
H^{(5)}_T&=&-\frac{1}{8 m^2} \int\!d{\bf x}\, \left( O_5 + O_6  - O_7
+ O_8  + 2\, O_{12} + O_{14} \right)  \ .
\end{eqnarray}

The constraints involving $H_2^{(5)}$ and the corresponding ${\bf W}_2^{(4)}$,
\begin{eqnarray}
\label{eq:cr2a}
\left [K^{(-1)i}_0\, ,\, W_2^{(4)j}\right] + 
\left[ W_2^{(4)i}\,  , \, K_0^{(-1)j} \right]&=&0 \ , \\
\label{eq:cr2b}
\left[{\bf K}^{(-1)}_0\, ,\, H_2^{(5)}\right]
+ \left[ {\bf W}_2^{(4)}\,  , \, H_0^{(0)} \right]&=& 0 \ ,
\end{eqnarray}
are fulfilled as long as the Hamiltonian $H^{(5)}_2$ is
constructed out of the 7 sub-leading operators listed
at the end of Sec.~\ref{sec:nr2}, or combinations thereof.
This is also shown in Appendix~\ref{app:a1}.  The
Hamiltonian $H^{(5)}_1 + H_2^{(5)}$
leads to the ${\bf P}$-dependent and
${\bf P}$-independent potentials in Eqs.~(\ref{eq:ct2})
and~(\ref{eq:ct2p}), in accordance with the derivation presented
in Sec.~\ref{sec:nr}.
 
In closing, we note that, although these constraints correspond to the
``minimal choice'' $\delta {\bf W}=0$, the result holds in the
general case.  Indeed, the requirement that $\delta {\bf W}$ commute with the
three-momentum operator implies that it be constructed as a spatial
integral of fields and their derivatives only.  No factors of
${\bf x}$, which would lower the counting power, are allowed inside the
integral.  The minimal power, in our counting, of an interacting (two-body)
boost operator $\delta {\bf W}$ is therefore 3, but is actually 4
if the relations~(\ref{eq:kkrel})--(\ref{eq:comm44a}) have to be
fulfilled order by order.  As a result, the only contributions of
$\delta {\bf W}$ to Eqs.~(\ref{eq:comm44}) and~(\ref{eq:comm44a}) are
given by $\left[\delta W^{(4)i}\, ,\, K_0^{(-1)j}\right]$ and
$\left[\delta {\bf W}^{(4)}\, ,\, H_0^{(0)}\right]$, both of which vanish,
since $\delta {\bf W}$, being hermitian, must contain
an equal number of creation and annihilation operators. 

\section*{Acknowledgments}
One of the authors (R.S.) would like to thank the Physics Department
of the University of Pisa, the INFN Pisa branch, and especially
the Pisa group for the support and warm hospitality extended
to him on several occasions.
The work of R.S.\ is supported by the U.S.~Department of Energy,
Office of Nuclear Physics, under contract DE-AC05-06OR23177.
\appendix
\section{Constraints on $H^{(3)}$ and $H^{(5)}$}
\label{app:a1}

In this appendix we outline the derivation of the leading- and next-to-leading order relations
in Eqs.~(\ref{eq:comm43})--(\ref{eq:comm43a}) and Eqs.~(\ref{eq:comm44})--(\ref{eq:comm44a})
involving $H^{(3)}$, as well as of the leading order relations in Eqs.~(\ref{eq:cr2a})--(\ref{eq:cr2b})
involving $H^{(5)}$. 
For brevity, we suppress spin indices, and introduce the notation
${\bm \nabla}_k\equiv {\bm \nabla}_{ {\bf p}_k}$, $b_k\equiv b_{s_k}({\bf p}_k)$, and
\begin{equation}
\int_{\bf p} \equiv \int
{\rm e}^{-i \left({\bf p}_1-{\bf p}_2+{\bf p}_3-{\bf p}_4\right)\cdot {\bf x}}
\, \prod_{k=1}^4\frac{d{\bf p}_k}{(2\pi)^3} \ .
\label{eq:adf1}
\end{equation}
Consider the commutator
\begin{equation}
\left[ K_0^{(-1)i}\, , \, W^{(2)j} \right]=\left[ K_0^{(-1)i}\, , \,
  C_S W_{S}^{(2)j}+ C_T W_{T}^{(2)j} \right] \ ,
\end{equation}
where the terms ${\bf W}_{S}^{(2)}$ and ${\bf W}_{T}^{(2)}$ correspond
to the interactions $H^{(3)}_S$ and $H^{(3)}_T$ [see Eq.~(\ref{eq:leadingH})],
for example
\begin{equation}
W_{S}^{(2)} = \frac{1}{2}\int d{\bf x}\, {\bf x}\, (N^\dagger N)(N^\dagger N)  \ .
\end{equation}
Making use of
\begin{equation}
\left[ {\bf K}_0^{(-1)}\, , \, b_1^\dagger\, b_2\, b_3^\dagger\, b_4  \right] =
-i\, m\, \left( \sum_{k=1}^4{\bm \nabla}_k\right)\,
b_1^\dagger\, b_2\,  b_3^\dagger\, b_4 \ ,
\end{equation}
we find that
\begin{equation}
 \left[ K_0^{(-1)i}\, , \, W_{S}^{(2)j} \right]=-i\, \frac{m}{2} \int d{\bf x}\,x^j\,  \int_{\bf p}
\left( \sum_{k=1}^4\nabla_k^{\,i} \right)\,
b_1^\dagger\, b_2\,  b_3^\dagger\, b_4 \ ,
\end{equation}
which vanishes after partial integrations with respect to the ${\bf p}_k$'s---note
the exponential factor in Eq.~(\ref{eq:adf1}).  The terms involving ${\bf W}^{(2)}_{T}$
as well as those occurring in $\left[ {\bf K}_0^{(-1)}\, , \, H^{(3)} \right]$ can
be worked out similarly, while those in $\left[ {\bf W}^{(2)}\, , \, H_0^{(0)} \right]$
vanish, since
\begin{equation}
\left[ b_1^\dagger\, b_2\,  b_3^\dagger\, b_4\, , \, H_0^{(0)} \right] = 0 \ .
\end{equation}
Thus each of the commutators entering the leading order relations
vanishes.

Moving on to the next-to-leading order relations, consider first $H_S^{(3)}$.  We obtain
\begin{equation} 
[{\bf K}^{(1)}_0\, , \, H_S^{(3)}] =  \frac{1}{2m}
\int d{\bf x} \int_{\bf p}  \left[ \big[ i\,  ({\bf p}_1 + {\bf p}_2 )
+ ( p_1^2 - p_2^2 )\, {\bf x}  \big] b_1^\dagger\, b_2\,  b_3^\dagger\, b_4 
+ b^\dagger_1\, \frac{{\bf p}_1 - {\bf p}_2}{2} \times {\bm \sigma}\,
b_2  \, b^\dagger_3 \,  b_4   \right] \ ,
\end{equation}
where the terms involving the momenta ${\bf p}_3$ and ${\bf p}_4$
reduce to those with ${\bf p}_1$ and ${\bf p}_2$ after
exchanging $3 \rightleftharpoons 1$ and $4 \rightleftharpoons 2$.
The linear term in ${\bf x}$ is canceled by
$\left[{\bf W}_{S}^{(2)}\, , \, H^{(2)}_0\right]$.  In order to cancel
the rest, one requires an interaction term $H_S^{(5)}$, given by 
\begin{eqnarray}
H_S^{(5)}&=& -\frac{1}{8m^2}\int d {\bf x} \int_{\bf p}
   \Big[ ( p_1^2 + p_2^2 + {\bf p}_2 \cdot {\bf p}_4 + {\bf p}_1\cdot
     {\bf p}_3 ) \, b^\dagger_1\,  b_2 \, b^\dagger_3\,  b_4
\nonumber \\
&& \, +i\,  b^\dagger_1\,  ({\bf p}_1 \times {\bf  p}_2 - {\bf p}_3 \times
{\bf p}_4 ) \cdot {\bm \sigma} \, b_2 \, 
b^\dagger_3 \,  b_4 \Big] \nonumber \\
&=& \frac{1}{8 m^2} \int d{\bf x} \Big[  ( N^\dagger\overleftarrow{\nabla}^2  N + N^\dagger
\overrightarrow{\nabla}^2 N) \,(N^\dagger N)  + ( N^\dagger \overrightarrow{{\bm
  \nabla}} N)^2 +  ( N^\dagger \overleftarrow{{\bm
  \nabla}} N)^2  \nonumber \\
&& \, + i\, ( N^\dagger
\overleftarrow{\bm \nabla} \cdot {\bm \sigma} \times  \overrightarrow{\bm \nabla}
 N ) \, (N^\dagger N) + i \, (N^\dagger {\bm \sigma} N) \cdot
(N^\dagger \overleftarrow{\bm \nabla} \times 
\overrightarrow{\bm \nabla} N)  \Big]\ ,
\end{eqnarray}
and a corresponding ${\bf W}_{S}^{(4)}$, which, however,
commutes with $H^{(0)}_0$.
Proceeding in a similar fashion for $H^{(3)}_T$, one finds that
an interaction term $H^{(5)}_T$,
\begin{eqnarray}
H_{T}^{(5)}\!\!&=&\!\! -\frac{1}{8m^2}\int d{\bf x} \int_{\bf p} 
 \Big[ \big[2\, \delta^{jk}\, {\bf p}_2 \cdot ({\bf p}_1 + {\bf p}_3)
     + 
p_2^j p_4^k - p_2^k p_4^j + p_1^j p_3^k - p_1^k p_3^j \big] \, b^\dagger_1\,  \sigma^j \, b_2 \,
b^\dagger_3\,  \sigma^k \,  b_4  \nonumber \\
&& \, 
+i\, b^\dagger_1\, ( {\bf p}_3 \times {\bf p}_4 - {\bf p}_1 \times
{\bf p}_2 ) \cdot {\bm \sigma} \,  b_2\,  b^\dagger_3\, b _4 \Big] \nonumber \\
&=&\!\!- \frac{1}{8m^2}\int d{\bf x} \Big[ - ( N^\dagger {\bm \sigma} \cdot
\overrightarrow{\bm \nabla} N) (N^\dagger {\bm \sigma}\cdot
\overrightarrow{\bm \nabla} N)  - ( N^\dagger {\bm \sigma} \cdot
\overleftarrow{\bm \nabla} N) (N^\dagger {\bm \sigma}\cdot
\overleftarrow{\bm \nabla} N) \nonumber \\
&& \, +(N^\dagger \sigma^j
\overrightarrow{\nabla^k} N)(N^\dagger \sigma^k
\overrightarrow{\nabla^j} N) + (N^\dagger \sigma^j
\overleftarrow{\nabla^k} N)(N^\dagger \sigma^k
\overleftarrow{\nabla^j} N) \nonumber \\
&& \, + 2 (N^\dagger \sigma^j \overrightarrow{\nabla^k} N) 
(N^\dagger \overleftarrow{\nabla^k} \sigma^j  N) +  2\, (N^\dagger
\overleftarrow{\bm \nabla} \sigma^j \cdot \overrightarrow{\bm \nabla}
N) (N^\dagger \sigma^j N) \nonumber \\
&& \, + i\, ( N^\dagger
\overleftarrow{\bm \nabla} \cdot {\bm \sigma} \times  \overrightarrow{\bm \nabla}
 N ) \, (N^\dagger N) + i \, (N^\dagger {\bm \sigma} N) \cdot
(N^\dagger \overleftarrow{\bm \nabla} \times 
\overrightarrow{\bm \nabla} N)  \Big] \ ,
\end{eqnarray}
and a corresponding boost operator ${\bf W}_T^{(4)}$, are required in order to satisfy the $T$-piece of
the commutators.  Thus Eq.~(\ref{eq:comm44a}) holds.
Similarly, Eq.~(\ref{eq:comm44}) can also be shown to hold.  The expressions for $H_S^{(5)}$
and $H^{(5)}_T$ correspond to those listed in Eqs.~(\ref{eq:h5s})--(\ref{eq:h5t}).

We now turn our attention to the constraint on $H_2^{(5)} $ implied by Eq.~(\ref{eq:cr2b}).
We first observe that the commutator $\left[ {\bf W}_2^{(4)}\,  , \, H_0^{(0)} \right]$
vanishes.  Defining
\begin{equation}
 \left[O_i\right] \equiv \int d{\bf x}\,  \left[ {\bf K}^{(-1)}_0, O_i\right]\ ,
\end{equation}
we find
{\allowdisplaybreaks 
\begin{align}
\left[  O_1\right] &= -2\,i\, m \int d{\bf x} \int_{\bf p} ({\bf p}_1  + {\bf p}_2)\, b_1^\dagger\, b_2\,  b_3^\dagger\, b_4\ ,  \\
\left[  O_2\right] &=      i\, m \int d{\bf x} \int_{\bf p} ({\bf p}_1  + {\bf p}_2)\, b_1^\dagger\, b_2\,  b_3^\dagger\, b_4\ ,  \\
\left[  O_3\right] &= -2\,i\, m \int d{\bf x} \int_{\bf p} ({\bf p}_1  + {\bf p}_2)\, b_1^\dagger\, b_2\,  b_3^\dagger\, b_4\ ,  \\
\left[  O_4\right] &=  0 \\
\left[  O_5\right] &= - m  \int d{\bf x} \int_{\bf p} ({\bf p}_1  - {\bf p}_2)\,
\times b^\dagger_1\, {\bm \sigma} \,b_2\, b^\dagger_3\, b_4 \ , \\
\left[  O_6\right] &=  m  \int d{\bf x} \int_{\bf p} ({\bf p}_3  - {\bf p}_4)\,
\times b^\dagger_1\, {\bm \sigma} \,b_2\, b^\dagger_3\, b_4\ , \\
\left[  O_7\right] &= -2\, i \, m  \int d{\bf x} \int_{\bf p} b^\dagger_1\,
{\bm \sigma}\, b_2 \, b^\dagger_3\, ({\bf p}_3  + {\bf p}_4) \cdot {\bm \sigma} \, b_4\ , \\
\left[  O_8\right] &= -2\, i \, m  \int d{\bf x} \int_{\bf p} b^\dagger_1\,
{\bm \sigma}\, b_2 \, b^\dagger_3\, ({\bf p}_1  + {\bf p}_2) \cdot {\bm \sigma} \, b_4\ , \\
\left[  O_9\right] &= -2\, i \, m  \int d{\bf x} \int_{\bf p}({\bf p}_1+{\bf p}_2)\,
b^\dagger_1{\bm \sigma}\,  b_2\cdot b^\dagger_3\, {\bm \sigma} \, b_4\ , \\
\left[  O_{10}\right] &=  i \, m  \int d{\bf x} \int_{\bf p}
b^\dagger_1{\bm \sigma}\,  b_2\, b^\dagger_3\, ({\bf p}_3+{\bf p}_4)\cdot{\bm \sigma} \, b_4\ ,\\
\left[  O_{11}\right] &=  i \, m  \int d{\bf x} \int_{\bf p}
b^\dagger_1{\bm \sigma}\,  b_2\, b^\dagger_3\, ({\bf p}_1+{\bf p}_2)\cdot{\bm \sigma} \, b_4 \ ,\\
\left[  O_{12}\right] &= i \, m  \int d{\bf x} \int_{\bf p}({\bf p}_1+{\bf p}_2)\,
b^\dagger_1{\bm \sigma}\,  b_2\cdot b^\dagger_3\, {\bm \sigma} \, b_4 \ , \\
\left[  O_{13}\right] &=  i \, m  \int d{\bf x} \int_{\bf p}
b^\dagger_1{\bm \sigma}\,  b_2\, b^\dagger_3\, ({\bf p}_1+{\bf p}_2+{\bf p}_3+{\bf p}_4)\cdot{\bm \sigma} \, b_4\ , \\
\left[  O_{14}\right] &=2\, i \, m  \int d{\bf x} \int_{\bf p}({\bf p}_1+{\bf p}_2)\,
b^\dagger_1{\bm \sigma}\,  b_2\cdot b^\dagger_3\, {\bm \sigma} \, b_4 \ ,
\end{align}}

\noindent and only 7 combinations of these operators satisfy the
constraint in Eq.~(\ref{eq:cr2b}), such as those in Eq.~(\ref{eq:7op}).
It is possible to show that these 7 combinations also satisfy the
constraint of Eq.~(\ref{eq:cr2a}).
\end{document}